\documentclass{ws-procs975x65}%
\usepackage{amsmath}
\usepackage{amsfonts}
\usepackage{amssymb}
\usepackage{graphicx}%
\begin{document}

\title{Cosmological Solutions in Biconnection and Bimetric Gravity Theories}

\author{Sergiu I. Vacaru}

\address{Alexandru Ioan Cuza University,
  14 Alexandru Lapu\c sneanu street,  Corpus R, UAIC, office 323;
Ia\c si,\  700057,\  Romania
\email{E--mails:\ sergiu.vacaru@uaic.ro, Sergiu.Vacaru@gmail.com}}

\begin{abstract}
We show how generic off--diagonal cosmological solutions depending, in general, on all spacetime coordinates can be constructed in massive gravity using the anholonomic frame deformation method. Such metrics describe the late time acceleration due to effective cosmological terms induced by nonlinear off--diagonal interactions and graviton mass and  include matter, graviton mass and other effective sources modelling nonlinear gravitational and matter fields interactions with polarization of physical constants and deformations of metrics, which may explain certain dark energy and dark matter effects.
\end{abstract}

\bigskip\bodymatter

Recently, a substantial progress in quantum gravity was made when de Rham and co--authors had shown how to eliminate the scalar mode and Hassan and Rosen established a complete proof for a class of bigravity / bimetric gravity theories \cite{drg}. The second metric describes an effective exotic matter related to massive gravitons and does not suffer from ghost instability to all orders in a perturbation theory and away from the decoupling limit.

We study modified massive gravity theories constructed on a pseudo--Riemannian spacetime $\mathbf{V}$ with physical metric $\mathbf{g}=\{\mathbf{g}_{\mu \nu }\}$ and  fiducial metric as we shall explain below. The action for our model is
{\small
\begin{equation}
S =\frac{1}{16\pi }\int \delta u^{4}\sqrt{|\mathbf{g}_{\alpha \beta }|}[\widehat{f}(\widehat{R})-\frac{\mathring{\mu}^{2}}{4}\mathcal{U}(\mathbf{g}_{\mu \nu },\mathbf{K}_{\alpha \beta })+\ ^{m}L] =\frac{1}{16\pi }\int \delta u^{4}\sqrt{|\mathbf{g}_{\alpha \beta }|}[f(R)+\ ^{m}L],  \label{act1}
\end{equation}%
}
where $\mathring{\mu}=const$ is the mass of graviton, $\widehat{R}$ is the scalar curvature for an "auxiliary" (canonical) connection $\widehat{\mathbf{D}}$ uniquely determined by $\mathbf{g}$ via conditions\cite{vadm1} 1) $\widehat{\mathbf{D}}\mathbf{g}=0,$ and 2) its $h$- and $v$-torsions are zero for the nonholonomically induced torsion $\widehat{\mathcal{T}}$ completely determined by
$\mathbf{g}$ for a conventional splitting $\mathbf{N}:T\mathbf{V=}h\mathbf{V}\oplus v\mathbf{V}$.  The "priority" of $\widehat{\mathbf{D}}$ is that it allows to
decouple the field equations in various gravity theories and construct exact solutions in very general forms.  We can impose additional nonholonomic (non--integrable constraints) when $\widehat{\mathbf{D}}_{\mid \widehat{\mathcal{T}}=0}\rightarrow \nabla $ and $\widehat{R}\rightarrow R,$ where $R$ is the scalar curvature of the Levi--Civita (LC) connection, $\nabla ,$ and it is possible to extract exact solutions in GR and various torsionless modifications.  We use the units when $\hbar =c=1$ and the Planck mass $M_{Pl}$ is defined via $M_{Pl}^{2}=1/8\pi G$ with 4--d Newton constant $G.$

The equations of motion for  modified massive gravity theory, MGT, (\ref{act1}) are
\begin{equation}
(\partial _{\widehat{R}}\widehat{f})\widehat{\mathbf{R}}_{\mu \nu }-\frac{1}{%
2}\widehat{f}(\widehat{R})\mathbf{g}_{\mu \nu }+\mathring{\mu}^{2}\mathbf{X}%
_{\mu \nu }=M_{Pl}^{-2}\mathbf{T}_{\mu \nu },  \label{mgrfe}
\end{equation}%
where  $\widehat{\mathbf{R}}_{\mu \nu }$ is the Einstein tensor for  $\mathbf{g}_{\mu \nu }$ and $\widehat{\mathbf{D}}$ and $\mathbf{T}_{\mu \nu }$ is the standard matter  energy--momentum tensor. \ For $\widehat{\mathbf{D}}\rightarrow \nabla ,$ we get $\widehat{\mathbf{R}}_{\mu \nu }\rightarrow R_{\mu \nu }$ with a standard Ricci tensor $R_{\mu \nu }$ for $\nabla .$ The effective energy--momentum tensor $\mathbf{X}_{\mu \nu }$ is defined in a "sophisticate" form by the potential of graviton
$\mathcal{U}=\mathcal{U}_{2}+\alpha _{3}\mathcal{U}_{3}+\alpha _{4}\mathcal{U}_{4}$, where $\alpha _{3}$ and $\alpha _{4}$ are free parameters. The values $\mathcal{U}_{2},%
\mathcal{U}_{3}$ and $\mathcal{U}_{4}$ are certain polynomials on traces of some other polynomials of a matrix $\mathcal{K}_{\mu }^{\nu }=\delta _{\mu
}^{\nu }-\left( \sqrt{g^{-1}\Sigma }\right) _{\mu }^{\nu }$ for a tensor determined by four St\"{u}ckelberg fields $\phi ^{\underline{\mu }}$ as  $\Sigma _{\mu \nu }=\partial _{\mu }\phi ^{\underline{\mu }}\partial _{\nu}\phi ^{\underline{\nu }}\eta _{\underline{\mu }\underline{\nu }}$,  when $\eta _{\underline{\mu }\underline{\nu }}=(1,1,1,-1).$ Following a
series of arguments \cite{kobayashi} we can  avoiding potential ghost instabilities if we fix  $\mathbf{X}_{\mu \nu }=\alpha ^{-1}\mathbf{g}_{\mu \nu }$.

De Sitter solutions  are possible, for instance, for ansatz of PG type\cite{kobayashi}
\begin{equation}
ds^{2}=U^{2}(r,t)[dr+\epsilon \sqrt{f(r,t)}dt]^{2}+\widetilde{\alpha }%
^{2}r^{2}(d\theta ^{2}+\sin ^{2}\theta d\varphi ^{2})-V^{2}(r,t)dt^{2}.
\label{pgm}
\end{equation}%
In above formula, there are used spherical coordinates labelled in the form $u^{\beta }=(x^{1}=r,x^{2}=\theta ,y^{3}=\varphi ,y^{4}=t),$ the function $f$\ takes non--negative values and the constant $\widetilde{\alpha }=\alpha /(\alpha +1)$ and $\epsilon =\pm 1.$ For such bimetric configurations, the St\"{u}ckelberg fields are parameterized in the unitary gauge as $\phi ^{\underline{4}}=t$ and $\phi ^{\underline{1}}=r\widehat{n}^{\underline{1}},\phi ^{\underline{2}}=r\widehat{n}^{\underline{2}},\phi ^{\underline{3}}=r \widehat{n}^{\underline{3}},$ where a three dimensional (3--d) unit vector is defined as $\widehat{n}=(\widehat{n}^{\underline{1}}=\sin \theta \cos \varphi ,\widehat{n}^{\underline{2}}=\sin \theta \sin \varphi ,\widehat{n}^{\underline{3}}=\cos \theta ).$ Any PG metric  (\ref{pgm}) defines solutions both in GR and in MGT. It allows us to extract the de Sitter solution, in the absence of matter, and to obtain standard cosmological equations with FLRW metric, for a perfect fluid source %
 $T_{\mu \nu }=\left[ \rho (t)+p(t)\right] u_{\mu }u_{\nu }+p(t)g_{\mu \nu }$, where $u_{\mu }=(0,0,0,-V)$ can be reproduced for the effective cosmological constant $\ ^{eff}\lambda =\mathring{\mu}^{2}/\alpha .$ It is also possible to express metrics of type (\ref{pgm}) in a familiar cosmological FLRW form (see formulas (23), (24) and (27) in first reference \cite{kobayashi}).

Let us consider an ansatz
\begin{eqnarray}
ds^{2} &=&\eta _{1}(r,\theta )\mathring{g}_{1}(r)dr^{2}+\eta _{2}(r,\theta ) \mathring{g}_{2}(r)d\theta ^{2} +\omega ^{2}(r,\theta ,\varphi ,t)\{\eta _{3}(r,\theta ,t)\mathring{h}%
_{3}(r,\theta )  \label{offdans} \\ && [d\varphi +n_{i}(r,\theta )dx^{i}]^{2}   +\eta _{4}(r,\theta ,t)\mathring{h}_{4}(r,\theta ,t)
[dt+(w_{i}(x^{k},t)+ \mathring{w}_{i}(x^{k}))dx^{i}]^{2}\},   \notag
\end{eqnarray}%
with Killing symmetry on $\partial _{3}=\partial _{\varphi },$ which (in general) can not be diagonalized by coordinate transforms. The values $\eta _{\alpha }$ are called "polarization" functions; $\omega $ is the so--called "vertical", v, conformal factor. The off--diagonal, N--coefficients, are labelled $N_{i}^{a}(x^{k},y^{4}),$ where (for this ansatz) $N_{i}^{3}=n_{i}(r,\theta )$ and $N_{i}^{4}=w_{i}(x^{k},t)+\mathring{w} _{i}(x^{k}).$ The  data for the "primary" metric are
 $\mathring{g}_{1}(r) = U^{2}-\mathring{h}_{4}(\mathring{w}_{1})^{2}, \mathring{g}_{2}(r)=\widetilde{\alpha }^{2}r^{2},\mathring{h}_{3}=\widetilde{%
\alpha }^{2}r^{2}\sin ^{2}\theta ,\mathring{h}_{4}=\sqrt{|fU^{2}-V^{2}|}, \mathring{w}_{1} =\epsilon \sqrt{f}U^{2}/\mathring{h}_{4},\mathring{w}%
_{2}=0,\mathring{n}_{i}=0$,  when the coordinate system is such way fixed that the values $f,U,V$ in (\ref{pgm}) result in a coefficient $\mathring{g}_{1}$ depending only on $r$.

We shall work with respect to a class of N--adapted (dual) bases
 $\mathbf{e}_{\alpha } = (\mathbf{e}_{i}=\partial _{i}-N_{i}^{b}\partial _{b},e_{a}=\partial _{a}=\partial /\partial y^{a})$ and $  
\mathbf{e}^{\beta } = (e^{j}=dx^{i},\mathbf{e}^{b}=dy^{b}+N_{c}^{b}dy^{c})$, 
which are nonholonomic (equivalently, anholonomic) because, in general, there are satisfied relations of type $\mathbf{e}_{\alpha }\mathbf{e}_{\beta
}-\mathbf{e}_{\beta }\mathbf{e}_{\alpha }=W_{\alpha \beta }^{\gamma }\mathbf{e}_{\gamma },$ for certain nontrivial anholonomy coefficients $W_{\alpha
\beta }^{\gamma }(u).$ For simplicity, we shall consider energy momentum sources  and effective cosmological constants which up to frame/coordinate transforms can be parameterized in the form  $\Upsilon _{\beta }^{\alpha }=\frac{1}{M_{Pl}^{2}(\partial _{\widehat{R}} \widehat{f})}(\mathbf{T}_{\beta }^{\alpha }+\alpha ^{-1}\mathbf{X}_{\beta
}^{\alpha })=\frac{1}{M_{Pl}^{2}(\partial _{\widehat{R}}\widehat{f})}(\ ^{m}T+\alpha ^{-1})\delta _{\beta }^{\alpha }=(\ ^{m}\Upsilon +\ ^{\alpha
}\Upsilon )\delta _{\beta }^{\alpha }$  for constant values $\ ^{m}\Upsilon$ and $\ ^{\alpha }\Upsilon$ with respect to N--adapted frames.

The system of field equations (\ref{mgrfe}) can be integrated in certain generic off--diagonal general forms depending on all spacetime coordinates \cite{vadm1},
{\small
\begin{eqnarray}
ds^{2} &=& e^{\psi (x^{k})}[(dx^{1})^{2}+(dx^{2})^{2}]+\frac{\Phi ^{2}\omega ^{2}}{4\ (\ ^{m}\Upsilon +\ ^{\alpha }\Upsilon )}\mathring{h}_{3}[d\varphi
+\left( \partial _{k}\ n\right) dx^{k}]^{2} \notag \\ 
 && -\frac{(\Phi ^{\ast })^{2}\omega ^{2}}{(\ ^{m}\Upsilon +\ ^{\alpha }\Upsilon )\Phi ^{2}}\mathring{h}_{4}[dt+(\partial _{i}\ \widetilde{A}) dx^{i}]^{2}.  \label{nvlcmgs}
\end{eqnarray}%
}
for any $\Phi =\check{\Phi},$ $(\partial _{i}\check{\Phi})^{\ast }=\partial _{i}\check{\Phi}^{\ast }$ and $w_{i}+\mathring{w}_{i}=\partial _{i}\check{%
\Phi}/\check{\Phi}^{\ast }=\partial _{i}\ \widetilde{A}.$ We can consider arbitrary nontrivial sources, $\ ^{m}\Upsilon +\ ^{\alpha }\Upsilon \neq 0,$ and generating functions, $\Phi (x^{k},t):=e^{\phi }$ and $n_{k}=\partial _{k}n(x^{i}).$  The polarization $\eta $--functions for (\ref{nvlcmgs}) are computed in the form
 $\eta _{1}=e^{\psi }/\mathring{g}_{1},\eta _{2}=e^{\psi }/\mathring{g} _{2},\eta _{3}=\Phi ^{2}/4\ (\ ^{m}\Upsilon +\ ^{\alpha }\Upsilon ),\eta
_{4}=(\Phi ^{\ast })^{2}/(\ ^{m}\Upsilon +\ ^{\alpha }\Upsilon )\Phi ^{2}$. 

It is possible to provide an "alternative" treatment of (\ref{nvlcmgs}) as exact solutions in MGT. We have to define and analyze the properties of fiducial St\"{u}ckelberg fields $\phi ^{\underline{\mu }}$ and the corresponding bimetric structure resulting in target solutions $\mathbf{g}=(g_{i},h_{a},N_{j}^{a})$: Let us analyze the primary
configurations related to $\mathring{\phi}^{\underline{\mu }}=(\mathring{\phi}^{\underline{i}}=a(\tau )\rho \widetilde{\alpha }^{-1} \widehat{n}^{\underline{i}},\mathring{\phi}^{\underline{3}}=a(\tau )\rho \widetilde{\alpha }^{-1}\widehat{n}^{\underline{3}},\mathring{\phi}^{\underline{4}%
}=\tau \kappa ^{-1}),$ when the corresponding prime PG--metric $\mathbf{\mathring{g}}$ is taken in FLRW form $ds^{2}=a^{2}(d\rho ^{2}/(1-K\rho
^{2})+\rho ^{2}(d\theta ^{2}+\sin ^{2}\theta d\varphi ^{2}))-d\tau ^{2}.$ The related fiducial tensor  is computed
 $\mathring{\Sigma}_{\underline{\mu }\underline{\nu }}du^{\underline{\mu }}du^{%
\underline{\nu }}=\frac{a^{2}}{\tilde{\alpha}^{2}}[d\rho ^{2}+\rho
^{2}(d\theta ^{2}+\sin ^{2}\theta d\varphi ^{2})+2H\rho d\rho d\tau -(\frac{%
\tilde{\alpha}^{2}}{\kappa ^{2}a^{2}}-H^{2}\rho ^{2})d\tau ^{2}]$, 
where the coefficients and coordinates are re--defined in the form $r\rightarrow \rho =\widetilde{\alpha }r/a(\tau )$ and $t\rightarrow \tau=\kappa t,$ for $K=0,\pm 1;\kappa $ is an integration constant; $H:=d\ln a/d\tau $ and the local coordinates are parameterized in the form $x^{\underline{1}}=\rho ,x^{\underline{2}}=\theta ,y^{\underline{3}}=\varphi ,y^{%
\underline{4}}=\tau .$

In summary, we have found new cosmological off--diagonal solutions in massive gravity with flat, open and closed spatial geometries. We applied a geometric techniques for decoupling the field equations and constructing exact solutions in $f(R)$ gravity, theories with nontrivial torsion and noholonomic constraints to GR and possible extensions  on (co) tangent Lorentz bundles. A very important property of such generalized classes of solutions is that they depend, in principle, on all spacetime coordinates via generating and integration functions and constants. After some classes of solutions were constructed in general forms, we can impose at the end nonholonomic constraints, cosmological approximations, extract configurations with a prescribed spacetime symmetry, consider asymptotic conditions etc. 

{\bf Acknowledgments:} Authors's research is partially supported by the Program IDEI, PN-II-ID-PCE-2011-3-0256.

\end{document}